\documentclass[12pt]{iopart}
\usepackage{multirow}
\usepackage{bbold}
\usepackage{subfigure}
\usepackage{color}
\usepackage{mathrsfs}
\usepackage{hyperref}
\usepackage[normalem]{ulem}
\usepackage{bm}

\usepackage{amsfonts, relsize, color}
\usepackage{graphics}
\usepackage{graphicx}
\usepackage{subfigure}
\usepackage{hyperref}
\usepackage{color}
\usepackage{comment}
\usepackage{graphicx,color,bm}
\usepackage[english]{babel}
\usepackage{epsfig}
\begin{document}

 \title{Thermal Magnetic Fluctuations of a Ferroelectric Quantum Critical Point }

 \author{Alexander Khaetskii$^1$, Vladimir Juri\v ci\' c$^2$, and Alexander V. Balatsky$^{1,2}$}
\address{Department of Physics, University of Connecticut, Storrs, CT 06269, USA}
\address{Nordita, KTH Royal Institute of Technology and Stockholm University, Roslagstullsbacken 23, SE-106 91 Stockholm, Sweden}


\begin{abstract}
Entanglement of two different quantum orders is of an interest of the modern condensed matter physics.
One of  the examples is the dynamical multiferroicity, where fluctuations of electric dipoles lead to magnetization.
 We investigate this effect at finite temperature and  demonstrate an elevated magnetic response of a ferroelectric near the ferroelectric quantum critical point (FE QCP).
We calculate the magnetic susceptibility of a bulk sample on the paraelectric side of the FE QCP at finite temperature and find  enhanced magnetic  susceptibility near the FE QCP.
We propose quantum paraelectric strontium titanate (STO) as a candidate material to search for dynamic multiferroicity.
We estimate the magnitude of the magnetic susceptibility  for this material and find that it is detectable experimentally.
\end{abstract}

\maketitle

Materials that can undergo phase transitions, especially those exhibiting  an interplay of two or more quantum orders attract special interest. One of  the examples featuring an entanglement of two different quantum orders is the dynamic multiferroicity (DMF), the concept reflecting the connection between electric and magnetic properties  \cite{Balatsky,Rebane,Dzyaloshinskii,Dunnett}.

Ferroelectric (FE) QCPs are an important part in  the discussion of FE behavior, particularly in displacive
quantum paraelectrics (PE)\cite{Khmelnitskii,Rowley,Chandra}. The plethora of effects that occur near a FE QCP have been investigated in various contexts \cite{Khmelnitskii,Rowley,Chandra,Rechester,Millis,Sondhi,Sachdev,Edge,Rischau,Narayan,Arce}.
In the DMF phenomenon magnetization ${\bf M}$ (i.e. magnetic moment per unit volume)  is induced by time-dependent oscillations of electrical polarization ${\bf P}$ (an electric dipole moment per unit volume):
\begin{equation}
{\bf M}=\lambda {\bf P}\times \frac{\partial {\bf P}}{\partial t}; \,\,\, \lambda \simeq \frac{a_0^3}{e_+ c}\frac{(m_- -m_+)}{(m_++m_-)}.
\label{magnetization}
\end{equation}
Here $a_0$ is the lattice spacing, and $c$ is the speed of light. The form of magnetization in the above equation is a consequence of the relation between the magnetic moment and the angular momentum for a rotating electric dipole, see also Eq.~\ref{eq:magneticmoment} and the discussion therein.   An expression for the coupling constant $\lambda$ is
obtained (Eq.~\ref{eq:magneticmoment}) for the model of optical long wave oscillations in the material with one cation and one anyon  per unit cell, with charges $\pm e_+$  and masses $m_\pm$. One can induce DMF by applying external drive-like light field to ferroelectrics. An alternative approach is to rely on inherent fluctuations present in the material near phase transition.

In this work we consider a bulk material in the vicinity of a ferroelectric phase transition, and investigate the enhanced magnetic susceptibility  at finite temperatures upon approaching the quantum critical point (QCP).
The main result of our work is the elevated magnetic response of a ferroelectric near the FE QCP, see Eq. (\ref{estimation}) and Fig. 1.
 We  calculate the dynamic magnetic susceptibility at finite temperature using the Kubo formula and  show that it grows in the paraelectric  phase upon approaching to the QCP,  which is controlled by the tuning parameter $\delta$ of the ferroelectric phase.  Material develops a paramagnetic susceptibility. Our estimates taken from possible candidate materials, such as SrTiO$_3$ (STO), show that the effect can be readily observed experimentally.

To understand the meaning of the result presented in Eq. (\ref{estimation}),
it is important to recall that the ferroelectricity is often induced by  the {\it softening} of the phonon optical mode, e.g. the frequency of this {\it ferroelectric} mode can gradually decrease upon lowering the temperature \cite{Yamada,Burke,Uwe}. In case of STO, the material we choose to illustrate the effect, FE phase transition occurs in the presence of strain or in the isotope-exchanged SrTi$^{18}$O$_3$, when the frequency of a ferroelectric mode reaches zero at finite temperature $T$ \cite{Itoh,Wang,Takesada,Taniguchi}. Based on these observations, we assume that near the FE QCP the  spectrum of a soft  transverse phonon mode reads as~\cite{Dzyaloshinskii,Millis}
\begin{equation}
\label{eq:phonon-spectrum}
\omega_{\perp}(q)=\sqrt{\omega_0^2(T,\delta)+s_{\perp}^2(\delta) q^2}.
\end{equation}
 Here, for the physically relevant low-energy phonon modes, by virtue of the general scaling arguments~\cite{Sondhi,Sachdev},
$ \omega_0(T,\delta)= \omega_0(T)|1-\delta|^{\nu z}$, $s_{\perp}(\delta)=s_0|1-\delta|^{\nu (z-1)}$,
 with $\omega_0(T)$ as a bare phonon gap, and $\nu$ ($z$) is the correlation length (dynamical) exponent.  We emphasize that the phonon gap and the transverse sound velocity may be gradually vanishing with the distance  $|1-\delta|$
 while approaching the QCP \cite{Rabe}, implying $z>1$ at the QCP  located at $\delta=\delta_c=1$.
Furthermore, higher order terms $\sim q^\alpha$, with $\alpha\geq4$, in the low-energy phonon dispersion  are even more irrelevant in the renormalization-group sense than the leading one proportional to the velocity, and as such can be neglected.

\par

For the (static) magnetic susceptibility in the high-$T$ limit, at $T\gg \hbar \omega_0(T,\delta)$, based on simple physical arguments (see below), supported by a formal analysis [see Eq. (\ref{Rezero})], we find
\begin{equation}
\chi (\omega=0) \simeq\left(\frac{e_+\hbar}{\bar{M}_0 c}\right)^2   \frac{T^2}{\hbar^3 s_{\perp}^3(\delta)}.
\label{estimation}
\end{equation}
Here $\bar{M}_0=m_+m_-/(m_++m_-)$, we have assumed that $(m_- -m_+)/(m_++m_-) \simeq 1$, and set $k_B=1$ hereafter.
Thus,  a strong increase of the magnetic susceptibility may occur while approaching the QCP:  the underlying reason for this enhancement is an increase of the density of states of phonons  due to the softening of the ferroelectric mode near the QCP, which is consistent with $z>1$.

We now provide the estimate of the magnetic susceptibility as given in Eq. (\ref{estimation}) using rather simple physical arguments. In our case, although the system is an insulator, we use the Pauli susceptibility because \emph{the fluctuating} electric dipoles, according to our mechanism, yield the magnetic response.
We start with the well known expression for the static Pauli susceptibility  $\chi\simeq \mu_B^2 \rho(\varepsilon) $, where $\mu_B$ is the Bohr magneton and $\rho(\varepsilon)$ is the density of states of the excitations at a corresponding energy. Then one needs to replace the Bohr magneton with the nuclear magneton $\mu_N$ (determined by the reduced mass of two atoms in a cell). Since phonon excitations yield the fluctuating dipoles which in turn give rise to the magnetization [see Eq.~(\ref{magnetization})], we calculate the density of states of phonons at a given temperature $T$, using the spectrum  given by Eq.~(\ref{eq:phonon-spectrum}). At $T\gg \hbar \omega_0$ the phonon density of states is $\rho(T) \simeq q^3_T/\hbar\omega_{\perp}(q_T) \simeq (1/\hbar s_{\perp})(T/\hbar s_{\perp})^2 $, where $q_T=T/\hbar s_{\perp}$ is the thermal momentum of a phonon, yielding  Eq.(\ref{estimation}).
The estimate in Eq.(\ref{estimation}) corresponds to the result obtained below more formally within the Kubo formalism, see Eqs.(\ref{ReQCP}) and (\ref{Rezero}).
The  case $T\ll \hbar \omega_0$  can be treated  in a similar manner \cite{low_T}.

 As mentioned before, one can reach the  vicinity of the critical value $\delta_c=1$ by applying strain or by  $O^{18}$ isotope substitution. For example, for the value of 95\% of isotope concentration, an approach to a critical point by changing temperature in a relatively narrow interval around $T_c \approx 25 K$ was achieved~\cite{Takesada}. Moreover,  the temperature dependence of the squared frequency of the ferroelectric mode near $T_c$ obeyed the Curie-Weiss law. This  might imply a suppression of the quantum fluctuations and our results obtained with the use of the bare Greens's functions of optical phonons can be applied in the close proximity of the critical point.

\begin{figure}[t!]
\includegraphics[width=1\linewidth]{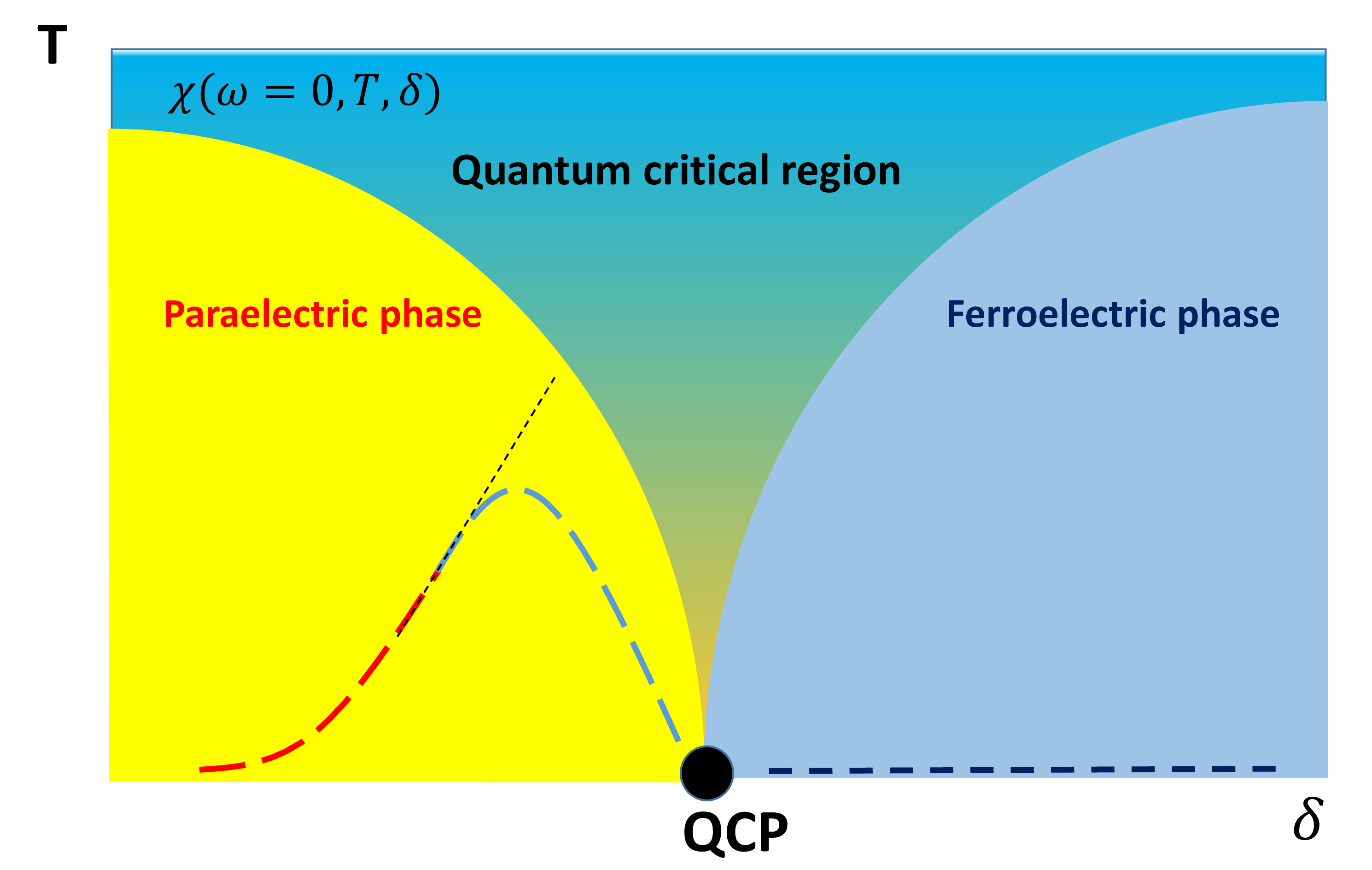}
\caption{Phase diagram near a ferroelectric QCP (where $\delta_c=1$). Static magnetic susceptibility in the paraelectric and ferroelectric phases  is shown by the dashed lines. Dotted line in the paraelectric phase corresponds to the behaviour described by Eq. (\ref{estimation}). A downturn  is expected in the magnetic susceptibility (dashed light blue line) upon decreasing the temperature and approaching the QCP. }~\label{Fig:phasediagram}
\end{figure}

To gain further insight into the role of the fluctuating dipoles associated with the soft phonon modes, it is instructive to derive an expression for the parameter $\lambda$ in Eq.~(\ref{magnetization}), which relates magnetization and electric polarization, by taking into account its microscopic origin. This analysis is based on the well-known relation between the magnetic moment ${\bm \mu}$ and the orbital angular momentum ${\bf L}$ for a point-like particle with electric charge $e$ and mass $m$, ${\bm\mu}=(e/2mc) {\bf L}$.
Considering the optical long wave oscillations in a system with two atoms per unit cell, we use $m_+{\bf u}_+ + m_-{\bf u}_-=0 $. Here
${\bf u}_+$ and ${\bf u}_-$ are the displacements of positive and negative ions.  Introducing relative displacement vector ${\bf u}={\bf u}_+ -{\bf u}_-$, we obtain for the orbital angular momentum of the unit  lattice cell,  ${\bf L}=(m_+m_-/(m_++m_-))[{\bf u}\times \dot{{\bf u}}]$, where $\dot{{\bf u}}\equiv \partial {\bf u}/\partial t$. For the magnetic moment of the unit cell ${\bf m}=(e_+/2c)
\sum_{\zeta=\pm}\zeta\,{\bf u}_\zeta\times \dot{{\bf u}}_\zeta$, we obtain
\begin{equation}\label{eq:magneticmoment}
{\bf m}=\frac{e_+}{2c}\frac{(m_- -m_+)}{(m_++m_-)}[{\bf u}\times \dot{{\bf u}}]=\frac{e_+(m_--m_+)}{2m_-m_+c}{\bf L},
\label{moment}
\end{equation}
where $e_+$ is the effective charge of the cation.
The expression for the coupling constant $\lambda$ directly follows from Eq.~(\ref{moment}), if one takes into account that ${\bf m}=a_0^3{\bf M}$ and ${\bf P}=e_+ {\bf u}/a_0^3$.

\par
{\it Magnetic susceptibility:  finite temperature case.-} We now derive our main results [including Eq.(\ref{estimation})] using the Kubo formalism at finite temperature. Note, that without an external magnetic field the average magnetization is zero,  $\langle{\bf M}\rangle=0$. We calculate the magnetic susceptibility $\chi_{ij}(\omega, {\bf k})$, as a function of (real) frequency $\omega$ and a wave vector ${\bf k}$ as given by the Kubo formula
\begin{equation}
\chi_{ij}(\omega, {\bf k})= \frac{i}{\hbar}\int_0^{\infty}dt \int d^3{\bf r}\, e^{i(\omega_+ t-{\bf k}{\bf r})}\langle[ M_i({\bf r}, t), M_j(0)]\rangle.
\label{Eq1}
\end{equation}
Here $\omega_+=\omega+i0$, $[,]$ is a commutator and $\langle...\rangle$ is the thermal average, with the magnetization in terms of the electric polarization as given by Eq. (\ref{magnetization}). Furthermore, the polarization created by the long wave optical phonons reads as \cite{Levinson}
\begin{equation}
P_i=\frac{1}{\sqrt{V}} \sum_{\xi,{\bf q}} A_q \gamma_{ij}h_{\xi,{\bf q}}^j [\hat {b}_{\xi,{\bf q}}  \,\, e^{i{\bf q}{\bf r}-i\omega_\xi t} +\hat {b}^\dagger_{\xi, {\bf q}} \,\,  e^{-i{\bf q}{\bf r} +i\omega_\xi t}],
 \label{Eq2}
\end{equation}
where $\gamma_{ij}$ is a piezotensor, ${\bf h}_{\xi,{\bf q}}$ is a unit vector of a phonon polarization, index $\xi$ labels
different phonon polarizations, $A_q=\sqrt{\hbar a_0^3/2\bar{M}_0\omega_\xi(q)}$, $\omega_\xi(q)$ is the phonon frequency, $a_0$ is the lattice spacing,  $\bar{M}_0$ is the reduced mass of the two atoms in the unit cell,
and $\hat b_{\xi,{\bf q}}$  ($\hat b^\dagger_{\xi,{\bf q}}$) are annihilation (creation) phonon operators. For simplicity, we consider a case when the tensor $\gamma_{ij}=\gamma \delta_{ij}$ is diagonal which corresponds to  a cubic  crystal system. Using Eq. (\ref{Eq1}) we then find
\begin{equation}
\chi_{ij}(\omega, {\bf k})=-\delta_{ij} \Lambda\sum_{\zeta=\pm} \int_{{\bf q}}
\frac{\zeta N[\omega_{\perp}(q)]}{\omega_+ +\zeta[\omega_{\perp}(q)-\omega_{\perp}(|{\bf q}-\zeta{\bf k}|)]},
\label{Final}
\end{equation}
where $\int_{\bf q}\equiv\int d^3{\bf q}/(2\pi)^3$,  $\omega_+=\omega+i 0$, and $\Lambda=(16/9)\lambda^2  \gamma^4 \hbar (a_0^3/\bar{M}_0)^2$. The details of the calculations are presented in the Supplemental Material (SM) \cite{Suppl}.
Note that the result is proportional to the Bose distribution function $N(\omega_{\perp})$. The zero temperature contribution is
discussed in the footnote \cite{note}.
By using expression for $\lambda$ from Eq. (\ref{magnetization}) and the  estimate
$\gamma \simeq e_+/a_0^3$ \cite{Levinson}, the quantity
$\Lambda$ can be expressed as   $\Lambda \simeq (e_+\hbar/\bar{M}_0 c)^2/\hbar \simeq \mu_N^2/\hbar$, where we have  assumed that $(m_- -m_+)/(m_++m_-) \simeq 1$. Therefore, the magnetic susceptibility in Eq.~(\ref{Final}) scales with the nuclear magneton.

\par
To illustrate the result in Eq.~(\ref{Final}), we first consider the case when the system is far from the QCP deep into the paraelectric phase, and the dispersion of the transverse optical phonons  is given by Eq.~(\ref{eq:phonon-spectrum}) with a fixed gap  $\omega_0$ and the transverse sound velocity $s_0$.
Moreover, we will assume that the temperature is low enough,  $T\ll \hbar \omega_0$, implying that the corresponding thermal momentum  $q_T \ll \omega_0/s_0$, with  $q_T \simeq \sqrt{T\omega_0}/(\hbar^{1/2} s_0)$. Expanding at small momenta $k\ll q_T$,  after performing the angular integration,   we obtain~\cite{Suppl}
\begin{equation}
\chi(\omega, k) =\Lambda \frac{ s_0^2}{\omega_0}\int_0^{\infty}  \frac{dq }{2\pi^2} \frac{q^2 N[\omega_{\perp}(q)]}{(s_0^2q/\omega_0)^2-((\omega+i0)/k)^2}.
\label{Final1}
\end{equation}
The denominator of the function in Eq.(\ref{Final1}) is the difference between a square of a phase velocity of the external probe, $\omega/k$, and a square of a group velocity of a phonon at a given momentum $q$,  $v_{{\rm gr}}(q)=s_0^2 q/\omega_0 $.  The characteristic group velocity of a phonon is $v_{gr}(T) \simeq s_0\sqrt{T/\hbar \omega_0}$, obtained at $q=q_T$.
Therefore,  considering the case $k\ll q_T$, $T\ll \hbar \omega_0$,   we find from Eq.~(\ref{Final1}) the imaginary part of the susceptibility in the form
\begin{equation}
\Im \chi(\omega, k)= \frac{\Lambda\omega_0^2 \omega}{4\pi  ks_0^4}
 e^{-\hbar \omega_0/T}e^{-(\omega/s_0k)^2(\hbar \omega_0/2T)},
\label{Im}
\end{equation}
while for the real part, up to corresponding higher order corrections, we obtain,
\begin{eqnarray}
\Re\chi(\omega, k) &=& \frac{\Lambda}{(2\pi)^{3/2}} \frac{1}{s_0^3}\sqrt{\frac{T\omega_0^3}{\hbar}} e^{-\frac{\hbar \omega_0}{T}},  \,\, \frac{\omega}{k} \ll v_{\rm gr}(T),
\label{dc} \\
\Re \chi(\omega, k)&=& -\frac{\Lambda}{s_0}\sqrt{\frac{T^3\omega_0}{(2\pi\hbar)^3}}  \left(\frac{k^2}{\omega^2}\right)e^{-\frac{\hbar \omega_0}{T}}, \,\, \frac{\omega}{k} \gg v_{\rm gr}(T).
\label{Re}
\end{eqnarray}
Note that $\Re\chi(0, k)$ coincides with the dc result of Dzyaloshinskii and Mills, see Eq.~(12c) in Ref.~\cite{Dzyaloshinskii}.

\par
{\it Vicinity of the QCP.-} We now consider the magnetic susceptibility  near the FE QCP where the  phonon spectrum assumes the  form as in Eq.~(\ref{eq:phonon-spectrum}).
Furthermore, in the close proximity of the QCP,  $|1-\delta| \ll 1$, and the gap is small comparing to the scale set by  temperature,  $\omega_0(T)(1-\delta)^{\nu z} \ll T$.   The main contribution to the  integral in Eq.~(\ref{Final}) then arises from the energies $\sim T$, when the spectrum $\omega_{\perp}(q)\approx s_{\perp}(\delta) q$ (we here neglect anomalous dimension of ferroelectric fluctuations, i.e. $\eta=0$).
At small momenta $k\ll q_T=T/\hbar s_{\perp}$, using  Eq.~(\ref{Final}), we then obtain
\begin{equation}
\Im \chi(\omega, k)\approx  \frac{\Lambda \pi \omega}{12  ks_{\perp}^2}
\left(\frac{T}{\hbar s_{\perp}}\right )^2\theta\left(1-\frac{|\omega|}{s_{\perp}k}
\right),
\label{Im1}
\end{equation}
where $\theta(x)$ is the step function. This result, together with the Kramers-Kronig relations, yields the following form of the real part of the susceptibility
\begin{equation}
\Re \chi(\omega, k) \approx \frac{\Lambda}{ 6s_{\perp}} \frac{T^2}{\hbar^2s_{\perp}^2} \left[1+\frac{\omega}{2ks_{\perp}}
     \ln \left|  \frac{ks_{\perp}-\omega}{ks_{\perp}+\omega}   \right|\right].
\label{ReQCP}
\end{equation}

Importantly, in the limit of the zero frequency, we then obtain
\begin{equation}
\Re \chi(0, k \to 0) \approx \frac{\Lambda}{6} \frac{T^2}{\hbar^2 s_0^3|1-\delta|^{3\nu (z-1)}},
\label{Rezero}
\end{equation}
with $\Lambda\simeq\mu_N^2/\hbar$, see  Eq.~(\ref{Final}). We therefore find  a strong increase of the susceptibility in the close proximity of the FE QCP due to the vanishing phonon velocity, $z>1$.

{\it Estimates and experimental proposal.-} Let us estimate the magnitude of the magnetic susceptibility, Eq.~(\ref{Rezero}), in the case of bulk SrTiO$_3$.  We consider first the case when the system  is far from the QCP, taking $s_{\perp}=s_0$. From the experimental data of Ref.~\cite{Yamada} one can estimate $s_0\approx 3\times 10^{5}$cm/s. Then taking $\bar{M}_0\approx 12m_p$, $e_+=4|e|$, and $T=50K$  we obtain $\chi(\omega=0) \simeq 10^{-11}$. If one takes $1-\delta \approx 0.1$, we obtain for the susceptibility near the QCP the  value  $\chi(\omega=0) \simeq 10^{-9}$. Both values are experimentally measurable.
\par
STO may be a suitable candidate material for the observation of magnetic signatures on tuning towards the FE QCP because of its incipient ferroelectric nature  \cite{Muller}.  Similarly, KTaO$_3$, due to its incipient ferroelectric features~\cite{perry}, is also expected to exhibit this phenomenology.
Several methods are used for tuning STO towards the FE QCP, such as  strain, applied pressure \cite{Burke,Uwe} or $^{18}$O substitution \cite{Itoh,Wang,Takesada,Taniguchi}.  Biaxial strain in
STO thin films, for example,  can confine polarization to the plane perpendicular to the tetragonal $c$ axis  but leaves undetermined the polarization direction. This creates  a favorable condition for the observation of the magnetic
signatures proposed here since it allows strong  fluctuations between different in-plane directions of the polarization.
\par
As for the method of tuning STO towards the FE QCP with the use of the oxygen isotope-exchange  samples, we should mention here two beautiful experiments \cite{Takesada,Taniguchi},  where  the authors demonstrated the ideal softening of the ferroelectric modes at  finite  temperature $T_c$.  To observe the effect proposed in this work, the
magnetic susceptibility should be measured as a  function of temperature in the close proximity of the phase transition point. A simple experimental setup consists of a SQUID placed above an STO sample.

{\it Conclusions.—} Multiferoicity as a static  phenomenon is well established.  On the other hand, the dynamically induced multiferroic phenonena are new and there has been no experiments that explicitly prove it. In this paper, we provide the theoretical estimates and rigorous bounds on the strength of the effect. Furthermore, we have expanded the framework of dynamic multiferroicity and predict strongly enhanced magnetic susceptibility in a paraelectric material near its FE QCP due to {\it thermal fluctuations} of the electric dipoles associated with a soft optical mode driving the transition.  The predicted effect indicates an alternative way for entangled quantum orders to be induced at finite temperature.  We thus further support the concept, proposed in Ref.~\cite{Dunnett}, that any FE QCP is an inherent  multiferroic QCP with
entangled ferroelectric and  ferromagnetic fluctuations.  Finally,  we have considered STO as a system that can be tuned towards its FE QCP, and provided an estimate of a magnetic  susceptibility for this material. The predicted value could be  experimentally accessible.  One of the possibilities to observe  magnetic signatures
of fluctuating dipoles is to make SQUID measurements.
\par
We are grateful to  K. Dunnett, N. Spaldin, I. Sochnikov, B. Spivak, J.-X. Zhu, C. P. Opeil and A. Polkovnikov   for useful discussions. The work was supported by VILLUM FONDEN via the Centre of Excellence for Dirac Materials (Grant No. 11744), Knut and Alice Wallenberg Foundation and the European Research Council ERC-2018-SyG HERO.
 V.J. acknowledges the support of the Swedish Research Council (VR 2019-04735).


\begin{thebibliography}{99}
\bibitem{low_T}
In the  case $T\ll \hbar \omega_0$ we should additionally take  into account the exponentially small  number of the excited phonons, $N \propto \exp(-\hbar \omega_0/T)$. In this case the density of states is $\rho(T)\simeq q_T^3/T \simeq T^{1/2}\omega_0^{3/2}/(\sqrt{\hbar}s_{\perp})^3$, where we use that $\omega_{\perp}(q_T)-\omega_0\approx (s_{\perp}^2q_T^2/2\omega_0)\simeq T/\hbar$. Then the product of  three terms, $\mu_N^2$, $\exp(-\hbar \omega_0/T)$, and $\rho(T)$, gives the result in agreement with the one obtained more formally, which is displayed in Eq.~(\ref{dc}).
\bibitem{note}
The zero temperature contribution can be estimated as $\Re \chi(0, k\to 0)\simeq \Lambda (k^2/s_{\perp}) \ln (1/q^*a_0)$, $q^*a_0\ll 1$. The characteristic phonon wave vector $q^* \simeq\omega_0(\delta)/s_{\perp}(\delta)$ is determined by the gap and the transverse sound velocity at a given distance from the QCP, given by the tuning parameter $\delta$.  Since we consider the limit $k\to 0$, we neglect this contribution. Notice that this zero temperature estimate is obtained with the use of the bare Greens' function for the soft phonons. In the quantum critical region the problem should be solved by accounting for the proper self-energy corrections to the phonon propagator and is the subject of the future investigation.
\end{thebibliography}
\end{document}